\documentclass[11pt,a4paper]{article}

\usepackage{physics}
\usepackage{empheq}
\usepackage{tensor}

\PassOptionsToPackage{dvipdfm}{hyperref}
\usepackage{jheppub}

\newcommand{\ifrac}[2]{{\left. #1 \middle/ #2 \right. }}
\newcommand{\siki}[1]{Eq.\eqref{#1}}

\newcommand{\Sec}[1]{Section \ref{#1}}
\renewcommand{\cal}[1]{\mathcal{#1}}
\newcommand{\ttb}{T\bar{T}}
\newcommand{\del}{\partial}
\newcommand{\app}[1]{Appendix \ref{#1}}

\DeclareMathOperator{\diag}{diag}

\newcommand{\vp}{\vec{\phi}}

\newcommand{\vpi}{\vec{\pi}}

\newcommand{\aln}[1]{\begin{align}#1\end{align}}

\begin{document}
\begin{flushright}
KUNS-2868
\end{flushright}

\title{Burgers Equation vs. Large $N$ Limit in $\ttb$-deformed $O(N)$ Vector Model}
\author[a]{Junichi Haruna,}  

\author[b]{Katsuta Sakai} 

\author[a]{and Kentaroh Yoshida}

\affiliation[a]{Department of Physics, Kyoto University, Kitashirakawa-Oiwakecho, Kyoto 606-8502, Japan}
\affiliation[b]{KEK Theory Center, High Energy Accelerator Research Organization (KEK), Oho 1-1, Tsukuba, Ibaraki 305-0801, Japan}

\emailAdd{j.haruna@gauge.scphys.kyoto-u.ac.jp}
\emailAdd{sakaika@post.kek.jp}
\emailAdd{kyoshida@gauge.scphys.kyoto-u.ac.jp} 

\abstract{We study a $\ttb$-deformed $O(N)$ vector model,  
which is classically equivalent to the Nambu-Goto action with static gauge.
The thermal free energy density can be computed exactly by using the Burgers equation as a special property of $\ttb$-deformation. The resulting expression is valid for an arbitrary value of $N$\,. One may consider a large $N$ limit while preserving this expression. We try to derive 
this result in the field-theoretical approach directly by employing the large $N$ limit.  
As a result, the leading contribution coincides with the exact one. That is, the $1/N$ corrections 
are cancelled out through a non-trivial mechanism.}

\maketitle

\section{Introduction}
Let us consider the theory space of 2D quantum field theories (QFTs), where the existence of the  Lagrangian description is assumed. One may consider a flow parameterized by a real constant parameter $\alpha$\,, 
along which the Lagrangian is denoted as $\mathcal{L}^{(\alpha)}$\,. 

\medskip

In the following, we will discuss a special flow caused by an irrelevant operator
\begin{eqnarray}
 \det\qty({T^{(\alpha)}_{~\mu\nu}}) \equiv  T^{(\alpha)}_{~00}T^{(\alpha)}_{~11} - (T^{(\alpha)}_{~01})^{2} \qquad \quad  (\mu,\nu=0,1)\,. 
\end{eqnarray}
This is just the determinant of the energy-momentum tensor derived from $\mathcal{L}^{(\alpha)}$ 
and often called the $\ttb$-operator. Its dimension is (length)$^2$ in natural units. Although this is a composite operator, it is well-defined as a local operator as shown by Sasha Zamolodchikov \cite{Zamolodchikov:2004ce}. 
Furthermore, the expectation value of this operator sandwiched between the excited states is factorized without taking some limits or relying on supersymmetry \cite{Zamolodchikov:2004ce}. 

\medskip

The flow caused by the $\ttb$-operator is called 
a $\ttb$-flow \cite{Smirnov:2016lqw,Cavaglia:2016oda}. 
The defining relation of the flow is given by 
\begin{eqnarray}
\mathcal{L^{(\alpha + \delta \alpha)}} = \mathcal{L^{(\alpha)}} 
+ \delta\alpha \det \qty({T^{(\alpha)}_{~\mu\nu}})\,. 
\label{relation}
\end{eqnarray}
This relation describes an infinitesimal deviation from $\mathcal{L}^{(\alpha)}$ (a deformed theory)\,, rather than $\mathcal{L}^{(0)}$ (the undeformed theory)\,.
It can be rewritten as a differential equation: 
\begin{align}
\pdv{\alpha}\cal{L}^{(\alpha)} = \det \qty(T^{(\alpha)}_{~\mu\nu}). 
\label{flow-eq}
\end{align}
This is called the $\ttb$-flow equation, 
and the deformation governed by it 
is called the $\ttb$-deformation. 

\medskip 

As an application of the $\ttb$-flow equation at classical level, we can derive a deformed Lagrangian $\mathcal{L}^{(\alpha)}$
by solving it directly. 
For example, if we consider the $\ttb$-deformation of 
a free massless scalar field theory, the deformed theory leads to the Nambu-Goto (NG) action (with static gauge) where the deformation parameter $\alpha$ is identified with $\alpha'$ in the string tension $T=1/(2\pi \alpha')$\,. 
Namely, the $\ttb$-deformation is closely related to string theory and provides a strong motive to study it. 

\medskip

A significant feature of the $\ttb$-deformation is that the energy spectrum of the deformed theory may be obtained explicitly.
If the system is defined on a infinitely-long cylinder, 
the deformation of the energy spectrum is governed 
by the inviscid Burgers equation 
\cite{Smirnov:2016lqw,Cavaglia:2016oda}: 
\begin{align}
\pdv{E_n^{(\alpha)}}{\alpha}
= E^{(\alpha)}_n\pdv{E_n^{(\alpha)}}{R}
+ R^{-1}\qty(P_n)^2\,, 
\label{B-eq}
\end{align}
where $E_n^{(\alpha)}$ and $P_n$ are the energy and momentum of the $n$-th excited state in the $\ttb$-deformed theory $\cal{L}^{(\alpha)}$ respectively, and $R$ is the circumference of the cylinder. If we know the energy spectrum of the undeformed theory $\mathcal{L}^{(0)}$ such as conformal field theory (CFT) and integrable quantum field theory (IQFT), we can get that of the deformed theory $\mathcal{L}^{(\alpha)}$ by solving the Buergers equation (\ref{B-eq}). 

\medskip

In a massless $O(N)$ vector model, for example, the deformed thermal free energy can be computed as 
\begin{eqnarray}
 f =\frac{1}{2\alpha}\qty(-1 +\sqrt{1-\frac{2\pi N}{3}\frac{\alpha}{\beta^2}}\,)\,,
\label{free-energy}
\end{eqnarray}
where $\beta\coloneqq T^{-1}$ is the inverse temperature.
This expression is valid for an arbitrary value of $N$\,. 
Note here that one may take a large $N$ limit \cite{Giveon:2017nie,Haruna:2020wjw}  
\begin{align}
 \alpha N = \tilde\alpha :\text{fixed}\,,\quad N\to \infty\,, 
\end{align}
preserving the expression of (\ref{free-energy})\,. It is significant 
to notice that this result is exact while it may be reproduced 
as well in the field-theoretical approach directly by employing the large $N$ limit. Our aim here is to try to carry out it. 

\medskip 

As another motive to consider the large $N$ limit, it is instructive to make some comments on our previous work \cite{Haruna:2020wjw}. 
It has been shown there that a $\ttb$-deformed free $O(N)$ vector model has negative-norm states inevitably (at least) in the large $N$ limit. Hence it would be interesting to see whether this negative norm state 
would be relevant to the Burgers result or not. 

\medskip

Based on the above motives, we compute the thermal free energy density in the field-theoretical approach by employing the large $N$ limit and compare the result to the exact one obtained by solving the Burgers equation\footnote{This large $N$ limit has also been discussed in \cite{Aharony:2018vux}, where $c=N$ is the total central charge. In this limit, correlation functions in $\ttb$-deformed CFTs may be computed exactly. For other aspects of correlation functions in $\ttb$-deformed theories, 
see \cite{Cardy:2019qao,Hirano:2020nwq,Hirano:2020ppu}.}.
As a result, we find that the leading contribution in the large $N$ limit coincides with the exact result. Our result indicates that the $1/N$ corrections 
are cancelled out through a non-trivial mechanism. 

\medskip

This note is organized as follows. 
Section \ref{s:rev} is a brief review of the $\ttb$-deformed massless $O(N)$ vector model and the derivation of its thermal free energy density through (\ref{B-eq}).
In section \ref{s:FNG}, we compute it in the large $N$ limit. We summarize and discuss the result in Section \ref{s:dis}.
Appendix \ref{app:CommentsOnHamiltonian} is the derivation of the Hamiltonian for the $\ttb$-deformed model. 
Appendix \ref{app:CalcVeff} is devoted to the detailed computation of the effective potential. 
In Appendix \ref{app:ExactBurgers}, we revisit the derivation of the Burgers equation in the operator formalism under some conditions.

\section{Review of the \texorpdfstring{$\ttb$}{TTbar}-deformed model}
\label{s:rev}
In this section, we shall introduce a free massless $O(N)$ vector model and its $\ttb$-deformation. 

\medskip

Our starting point is the classical Lagrangian in 2D Euclidean space with the 2D coordinates $(x^1,x^2)$. 
The undeformed Lagrangian is given by 
\begin{align}
\label{e:FreeMasslessScalar}
    \cal{L} &= \frac{1}{2}\delta^{\mu\nu}\,\del_\mu \vp \cdot \del_\nu \vp\,,
\end{align}
where each of $\phi^m~(m=1,\ldots,N)$ is a real scalar field and $\delta^{\mu\nu}=\diag(1,1)$ is the (inversed) 2D Euclidean metric.
By solving the flow equation (\ref{flow-eq})\,, the deformed Lagrangian \cite{Cavaglia:2016oda,Bonelli:2018kik} is explicitly obtained as   
\begin{multline}
\label{e:LN}
\cal{L}=
	\frac{1}{2\alpha}\biggl(-1+
	\sqrt{
		1 + 2\alpha\Bigl[\qty(\del_1\vec{\phi})^2+\qty(\del_2\vec{\phi})^2\Bigr]+4\alpha^2
		\Bigl[\qty(\del_1{\vec{\phi}})^2 \qty(\del_2\vec{\phi})^2-\qty(\del_1 \vec{\phi}\cdot \del_2 \vec{\phi})^2\Bigr]
		}\biggr)\,, 
\end{multline}
where $\alpha$ is a deformation parameter with dimension (length)$^2$\,.
This Lagrangian can be identified 
with the Nambu-Goto (NG) Lagrangian
\begin{align}
\cal{L}_{\rm NG}&= \sqrt{\det(h_{\mu\nu})}\,,
\end{align}
with static gauge (up to a constant term).
Here $h_{\mu\nu}$ is induced from the target space of the $(N+2)$-D Euclidean spacetime with the metric $\delta_{mn}$ and is defined as
\begin{align}
    h_{\mu\nu} & \coloneqq  \delta_{mn}\,\del_\mu \phi^m \del_\nu \phi^n
    \qquad (m,n =1,\ldots,N+2)\,. 
\end{align}
The static gauge is realized as
\begin{align}
    \phi^{\,N+1} = \frac{1}{\sqrt{2\alpha}}x^1\,,
    \qquad 
    \phi^{\,N+2} = \frac{1}{\sqrt{2\alpha}}x^2\,
\end{align}
and then $h_{\mu\nu}$ is evaluated as 
\begin{align}
    h_{\mu\nu}& = \frac{1}{2\alpha}\delta_{\mu\nu}+\del_\mu\vec{\phi}\cdot\del_\nu\vec{\phi}\,.
\end{align}
In this expression, the fields $\vec{\phi}$ describe classical fluctuations around 2D flat subspace in $(N+2)$-D target space. 

\medskip

The derivation of the thermal free energy density with the Burgers equation is the following. 
There are two points to note. First, with the inverse temperature $\beta=T^{-1}$, the thermal free energy density $f(\beta)$ on flat space is computed from the ground-state energy $E_0(R)$ 
on the cylinder with the circumference $R$ through the relation  \cite{Zamolodchikov:1989cf} 
\begin{align}
f(\beta)=\frac{E_0(R)}{R}\eval{}_{R=\beta}.
\label{e:rel_fe0}
\end{align}
Second, the $\ttb$-deformed energy spectrum on 
the cylinder can be computed by solving the Burgers equation (\ref{B-eq})\,.

\medskip 

For the undeformed theory (\ref{e:FreeMasslessScalar}) on the cylinder, the ground-state energy $E_0^{(\rm free)}$ is just the Casimir energy and is given by 
\begin{align}
    \frac{E_0^{(\rm free)}(R)}{R} = - \frac{\pi N}{6R^2}\,.
\end{align}
By taking it as the initial value to solve the Burgers equation (\ref{B-eq}), we obtain the ground-state energy 
of the $\ttb$-deformed theory \cite{Smirnov:2016lqw,Cavaglia:2016oda}: 
\begin{align}
\frac{E_0^\qty(\alpha)(R)}{R}=\frac{1}{2\alpha}\qty(-1 +\sqrt{1-\frac{2\pi N}{3}\frac{\alpha}{R^2}})\,.
\end{align}
Here we have used the fact that the total momentum of the ground state 
vanishes for any finite value of $\alpha$\,. 
Thus, the formula (\ref{e:rel_fe0}) leads to the thermal free energy density:  
\begin{align}
f 
&=\frac{N}{2\tilde{\alpha}}\qty(-1 +\sqrt{1-\frac{2\pi }{3}\frac{\tilde{\alpha}}{\beta^2}}\,)\,. 
\label{e:freeE_Burgers}
\end{align}
In the last equality, we have introduced a new quantity 
\begin{eqnarray}
\tilde{\alpha} \equiv N \alpha 
\end{eqnarray}
for later convenience in considering a large $N$ limit.

\section{Free energy density in large \texorpdfstring{$N$}{N}}
\label{s:FNG}

In this section, let us derive the thermal free energy density in the field-theoretical approach by employing a certain large $N$ limit.  

\subsection{Derivation}

We start from the following partition function\footnote{Note that we cannot start naively from the Lagrangian because the Lagrangian (\ref{e:LN}) includes higher-order terms of $\del_0\vp$ in the Lorentzian formulation. 
This point is also mentioned in a recent work \cite{He:2020cxp}.
}
\begin{align}
Z=\int D\vec{\phi}D\vec{\pi} \exp(\int_0^{\beta} dx^2 \int_{\mathbb{R}} dx^1\,\qty{i\vec{\pi}\cdot \del_2\vec{\phi}-\cal{H}}). 
\label{e:exactPathInt}
\end{align}
As usual, the fields have periodicity of $\beta$ in the imaginary-time direction $x^2$, and the model is essentially defined on the cylinder. The Hamiltonian is derived as\footnote{For the derivation of this Hamiltonian, see \app{app:CommentsOnHamiltonian}.}
\begin{align}
\label{e:HN}
\cal{H}
=\frac{1}{2\alpha}\qty[
-1 +
\sqrt{
		1 + 2\alpha\qty(\del_1\vec{\phi})^2 + 2\alpha\vec{\pi}^2 + 4\alpha^2\qty(\vec{\pi}\cdot\del_1\vec{\phi})^2
	}
]\,,
\end{align}
where $\vec{\pi}$ is the conjugate momentum of $\vp$\,.

\medskip

To evaluate the partition function, 
we consider here a large $N$ limit \cite{Giveon:2017nie,Haruna:2020wjw}:
\begin{align}
 \alpha N = \tilde\alpha ~:\text{fixed}\,,~~\quad N\rightarrow \infty\,. \label{largeN}
\end{align}
As shown later, $\vp$ and $\vpi$ can formally be integrated out by introducing auxiliary fields. Then, in the large $N$ limit (\ref{largeN}),  the integrals of the auxiliary fields can also be evaluated perturbatively with respect to $\ifrac{1}{N}$\,.

\medskip

Let us now calculate the thermal partition function.
In the following, we will consider the $\alpha>0$ case based on the physical ground. 
By rescaling the fields as
\begin{align}
\sqrt{2\alpha}\,\vec{\pi} ~\to~ \vec{\pi}\,, \qquad \sqrt{2\alpha}\,\vec{\phi} ~\to~ \vec{\phi}\,,  
\label{e:rescaling}
\end{align}
the partition function is rewritten as 
\begin{align}
Z
&=\int\!\! D\vec{\phi}D\vec{\pi}\, 
\exp(
	\frac{N}{2\tilde\alpha} \int\!\! d^2x 
	\left[
		i\vec{\pi}\cdot \del_2\vec{\phi} +1 -
			\sqrt{
						1 + \qty(\del_1\vec{\phi})^2 + \vec{\pi}^2+\qty(\vec{\pi}\cdot\del_1\vec{\phi})^2
					}\, 
					\right]
		)\,, 
\label{e:ZGeneralPathInt2}
\end{align}
where we have introduced a shorthand notation $\int \! d^2x$ for $\int_0^{\beta}\! dx^2 \int_{\mathbb{R}}\! dx^1 $.

\medskip

Then, let us introduce auxiliary fields $\lambda$ and $\rho$\,.
By inserting the identity
\begin{align}
    1&=\int\!\! D\rho\, \prod_{x}\delta \qty(\qty(\del_1\vec{\phi}(x))^2 + \vec{\pi}(x)^2 + \qty(\vec{\pi}(x)\cdot\del_1\vec{\phi}(x))^2-\rho(x))
    \notag  \\&= 
    \int \!\! D\lambda D\rho\,\exp\qty(
        \frac{iN}{2\tilde\alpha}\int d^2x\, \lambda
        \qty[
            \qty(\del_1\vec{\phi})^2 +\vec{\pi}^2 + \qty(\vec{\pi}\cdot\del_1\vec{\phi})^2 - \rho
        ]
        )
\label{e:IdDelta}
\end{align}
to the partition function (\ref{e:ZGeneralPathInt2})\,, 
we obtain the following expression: 
\begin{align}
\quad Z&=\int\!\! D\vec{\phi}D\vec{\pi} D\lambda D\rho\, \exp\biggl(
	\frac{N}{2\tilde\alpha} \int\!\! d^2x\,
		\biggl[ 
		i\vec{\pi}\cdot \del_2\vec{\phi} +1 -
			\sqrt{
						1 + \rho
					}
\notag\\
&\hspace{120pt}+i\lambda \Bigl[ \qty(\del_1\vec{\phi})^2 + \vec{\pi}^2 + \qty(\vec{\pi}\cdot\del_1\vec{\phi})^2-\rho\Bigr]
		\biggr]
\biggr)
\notag\\
&=\int\!\! D\vec{\phi}D\vec{\pi} D\lambda D\rho D\sigma 
\exp\biggl(
	\frac{N}{2\tilde\alpha} \int\!\! d^2x\,
		\biggl[
		i\vec{\pi}\cdot \del_2\vec{\phi} + 1 -
			\sqrt{
						1 + \rho
					}\notag\\
			&\hspace{100pt}+i\lambda \Bigl[ \qty(\del_1\vec{\phi})^2+\vec{\pi}^2 + 2\sigma \vec{\pi}\cdot\del_1\vec{\phi} - \sigma^2-\rho\Bigr]
		\biggr]+\frac{1}{2}\Tr \log \lambda
\biggr).
\label{e:PI_rw}
\end{align}
In the first equality, we have replaced $(\del_1\vec{\phi})^2 +\vec{\pi}^2+(\vec{\pi}\cdot\del_1\vec{\phi})^2$ 
with $\rho$ in the square root by using the delta function.
In the second equality, we have used the Gaussian integral formula 
\begin{align}
&  \exp\qty(\frac{iN}{2\tilde\alpha}\int\!\! d^2x\,\lambda\qty(\vec{\pi}\cdot\del_1\vec{\phi})^2)
\notag
    \\
   \propto & \int\!\! D\sigma\, \exp
    \qty(
        \frac{1}{2}\Tr\log\lambda + \frac{iN}{2\tilde\alpha}\int\!\! d^2x\, \lambda
        \qty(
            - \sigma^2 +2\sigma \vec{\pi}\cdot\del_1\vec{\phi}
            )
        )
\end{align}
to rewrite the four-point interaction of $(\vec{\pi}\cdot \del_1\vec{\phi})^2$. 
Note here that the symbol ``\,$\Tr$\,'' denotes the trace over the functional space. Since the argument of the exponential in (\ref{e:PI_rw}) is a quadratic form of $\vec{\pi},\vec{\phi}$, we can integrate out them. The resulting form is given by 
\begin{align}
Z
&= \int D\lambda D\rho D\sigma \exp(-S_{\rm eff})\, 
\end{align}
with the effective action $S_{\rm eff}$ for $\lambda$\,, $\rho$ and  $\sigma$ 
of the following form: 
\begin{multline}
\frac{S_{\rm eff}}{N} \coloneqq \frac{1}{2}\Tr \log \qty[\lambda \del_1 \lambda \del_1-\frac{1}{4}(\del_2 +2\del_1 \lambda \sigma )(\del_2 +2\lambda \sigma \del_1)]
\\  -
	\frac{1}{2\tilde\alpha} \int\!\! d^2x\,
		\qty( 1
			-\sqrt{
						1+\rho
					}
			-i\lambda \qty( \sigma^2 +\rho)
		)
 -\frac{i}{2N}\Tr \log \lambda\,.
\label{e:se_g}
\end{multline}
Note here that this expression is valid for finite values of $N$\,. 

\medskip

Then, let us consider the large $N$ limit (\ref{largeN}) and evaluate the leading-order contribution.
In this limit, the stationary point of $S_{\rm eff}$ dominates in the integrals over the auxiliary fields.
The stationary conditions are 
\begin{align}
\fdv{S_{\rm eff}}{\lambda(x)}=\fdv{S_{\rm eff}}{\rho(x)}=\fdv{S_{\rm eff}}{\sigma(x)}=0\,.
\end{align}
Because $\lambda$\,, $\rho$ and $\sigma$ are constant due to translational invariance of the system, 
it suffices to focus on the effective potential: 
\begin{align}
\frac{V_{\rm eff}}{N}&=
-\frac{\pi }{6\beta^2}\frac{1}{2\abs{\lambda}\abs{1 - \sigma^2}}
-
	\frac{1}{2\tilde\alpha} 
		\qty( 1-
			\sqrt{
						1+\rho
					}
			-i\lambda \qty( \sigma^2 +\rho)
		)
\,.
\label{e:Veff}
\end{align}
For the detailed derivation of (\ref{e:Veff}), see \app{app:CalcVeff}.
Note that the third term in the right-hand side of (\ref{e:se_g}) vanishes at the leading in large $N$\,.

\medskip

The stationary conditions for constant $\lambda$, $\rho$ and $\sigma$  
\begin{align}
\pdv{V_{\rm eff}}{\lambda}=\pdv{V_{\rm eff}}{\rho}=\pdv{V_{\rm eff}}{\rho}=0 
\end{align}
 are solved as
\begin{align}
i\lambda=-\frac{1}{2}\qty(1 - \frac{2\pi}{3}\frac{\tilde\alpha}{\beta^2})^{\frac{1}{2}}\,, 
\qquad \sigma=0,
\qquad \rho=\qty(1 - \qty(\frac{2\pi}{3}\frac{\tilde\alpha}{\beta^2})^{-1})^{-1}\,.
\end{align}
The value of the effective action at the stationary point is evaluated as 
\begin{align}
\frac{S_{\rm eff}}{N}\eval{}_{\text{stat pt, large }N}
&=\frac{1}{N}\int d^2xV_{\text{eff}}\eval{}_{\text{stat pt, large }N}\notag\\
&=\int d^2x\,\frac{1}{2\tilde{\alpha}}\qty(-1 +\sqrt{1 - \frac{2\pi}{3}\frac{\tilde\alpha}{\beta^2}})\,.
\end{align}
Then the leading contribution to the the free energy density is given by 
\begin{align}
f&= 
\frac{N}{2\tilde{\alpha}}\qty(-1 +\sqrt{1 - \frac{2\pi}{3}\frac{\tilde\alpha}{\beta^2}})
\label{e:freeE_NG}
\end{align}
and  {\it does} coincide with the exact result (\ref{e:freeE_Burgers}).

\subsection{Miscellaneous comments}

Let us make some comments on our result as follows. 

\subsubsection*{Negative \texorpdfstring{$\alpha$}{alpha0}}
So far, we have supposed that the deformation parameter $\alpha$ is positive. But even for the negative $\alpha$\,, one can derive the same results by following almost the same manner in the previous subsection. 

\medskip

When $\alpha$ is negative, we should rescale the fields as
\begin{align}
    \sqrt{-2\alpha}\vp \to \vp\,,\qquad \sqrt{-2\alpha}\vpi \to \vpi\,,
\end{align}
instead of \siki{e:rescaling}.
By using the following identity 
\begin{align}
    1= 
    \int D\lambda D\rho\,\exp\qty(
        \frac{iN}{2\tilde\alpha}\int d^2x\, \lambda\qty[
            -\qty(\del_1\vec{\phi})^2 - \vec{\pi}^2 + \qty(\vec{\pi}\cdot\del_1\vec{\phi})^2 - \rho
        ]
        )\,,
\end{align}
the resulting effective action is given by   
\begin{multline}
\frac{S_{\rm eff}}{N} \coloneqq \frac{1}{2}\Tr \log \qty[\lambda \del_1 \lambda \del_1-\frac{1}{4}(\del_2 - 2\del_1 \lambda \sigma )(\del_2 -2\lambda \sigma \del_1)]
\\  -
	\frac{1}{2\tilde{\alpha}} \int d^2x\,
		\qty( 1
			-\sqrt{
						1+\rho
					}
			-i\lambda \qty( \sigma^2 +\rho)
		)
 -\frac{i}{2N}\Tr \log \lambda\,.
\label{e:se_n}
\end{multline}
The difference from \siki{e:se_g} is just the sign of $\del_1$, which can be absorbed into the redefinition of $\sigma$. 
As a result, we get the same effective potential as in (\ref{e:Veff}).
Hence the leading contribution is valid even for the negative $\alpha$ as well.

\subsubsection*{Mircocanonical entropy}
The microcanonical entropy density is calculated by the Legendre transformation of the free energy density with respect to the temperature.
It is given by
\begin{align}
    s \coloneqq \min_{T} \qty(\frac{e-f}{T}) 
    = \sqrt{\frac{2\pi N}{3} e\qty(1+\frac{\tilde{\alpha}}{N}e)}
    = \sqrt{\frac{2\pi N}{3} e\qty(1+\alpha e)}\,,
    \label{e:MCEntropy}
\end{align}
where $e$ is the internal energy density.
In the high energy limit ($e\to \infty$), $s$ behaves as 
\begin{align}
    s \sim \sqrt{\frac{2\pi N}{3}\alpha}\,e = \beta_H\, e\,, \qquad 
    \beta_H = (T_H)^{-1} \coloneqq \qty(\frac{2\pi N\alpha}{3})^{-\frac{1}{2}}\,, 
\end{align}
where $T_H$ is the Hagedorn temperature \cite{Giveon:2017nie}.
This result can also be obtained from the Cardy formula \cite{Cardy:1986ie}.

\medskip 

Since the microcanonical entropy density counts the number of energy eigenstates with energy $e$\,, some information on the negative-norm states \cite{Haruna:2020wjw} may be encoded into it.
However, it seems likely that their contributions do not appear in (\ref{e:MCEntropy})\,. This issue is a puzzle to be resolved in the future work.

\subsubsection*{Derivation of Burgers equation}
As a possibility, one might suspect that the Burgers equation is not exact. 
In the original derivation \cite{Zamolodchikov:2004ce}, 
a physical interpretation of the expectation value of $T_{11}$ with excited states 
is used and might make a loophole. In \app{app:ExactBurgers}, we have proven that the Burgers equation is valid without using the physical interpretation by supposing that 1) the $\ttb$ operator is well-defined as a local operator, 2) the expectation value of the $\ttb$ operator is factorized and 3) the Hamiltonian consists only of mutually commutative operators.
The third condition is very strong and the situation becomes  restrictive. But the $\ttb$-deformed massless $O(N)$ vector model is included in this analysis and the argument there is a bit general. 
Hence, this possibility has been excluded in this class of theories.

\subsubsection*{UV divergences}
It is also helpful to make a comment on the ultraviolet divergences.
Recall that we have not explicitly encountered them in the calculation in \Sec{s:FNG}.
This is just because we have implicitly tuned the cosmological constant so that the free energy should be zero at the zero-temperature limit ($\beta\to\infty$).

\section{Conclusion and Discussion}
\label{s:dis}

In this note, we have discussed the thermal free energy density in a $\ttb$-deformed $O(N)$ vector model.
It can be computed by using the Burgers equation and the resulting expression is valid for an arbitrary value of $N$\,.
Since we may take a large $N$ limit so as to keep this expression, we have tried to derive this result in the field-theoretical approach directly by employing the large $N$ limit.  
As a result, the leading contribution coincides with the exact one. 

\medskip 

What would have happened about the sub-leading contributions to the large $N$ result (\ref{e:freeE_NG})?
This result indicates that the large $N$ calculation gives rise to the exact result and there should be a non-trivial cancellation mechanism in the sub-leading corrections.
The cancellation may come from the integrability of the $O(N)$ vector model preserved under the $\ttb$-deformation.
As another possibility, beyond the integrability, it may be related to the one-loop exactness discussed by Dubovsky et al \cite{Dubovsky:2018bmo} in the context of the Jackiw-Teitelboim (JT) gravity \cite{Teitelboim:1983ux,Jackiw:1984je}. 
It is a significant future problem to reveal this cancellation mechanism explicitly. 

\medskip

Although the massless $O(N)$ vector model has been studied so far, one may consider the massive case. Then, the undeformed theory is a free massive theory and there is no conformal symmetry (though the $\ttb$-deformed theory is still integrable). Hence the resulting entropy is not governed by the Cardy formula and it would enable us to see more apparently the relation between the thermodynamic property and integrability. 

\medskip 

One may also consider a generalization to higher-dimensional cases with a higher-dimensional version of $\ttb$-operator \cite{Bonelli:2018kik,Cardy:2018sdv,Taylor:2018xcy}.
It is curious to study whether the non-trivial cancellation would occur even in higher dimensions or not.
It is also nice to try to interpret our result from the bulk point of view in the context of AdS/CFT. 

\medskip 

We hope that our result would shed light on the mysterious property of the $\ttb$-deformation 
and find an interesting application in String Theory.

\section*{Acknowledgement}
We would like to thank Hikaru Kawai and Takaaki Ishii for useful discussions.
The work of K.S.\ is supported in part by the Grant-in-Aid for Scientific research, No. 20J00079.
The work of K.Y.\ is supported by the Supporting Program for Interaction-based Initiative Team Studies (SPIRITS) from Kyoto University, and JSPS Grant-in-Aid for Scientific Research (B) No.\,18H01214. 
This work is also supported in part by the JSPS Japan-Russia Research Cooperative Program.

\newpage
\appendix
\renewcommand{\theequation}{\Alph{section}.\arabic{equation}}

\section*{Appendix}

\section{Derivation of the Hamiltonian}
\label{app:CommentsOnHamiltonian}

In Section 3.1, we had to start from the Hamiltonian (\ref{e:HN}) in the path-integral formulation. But in the main text of the present note, we have entirely worked in the Euclidean signature, while in deriving the Hamiltonian (\ref{e:HN}) we need to begin in the Lorentzian formulation so as to follow the standard manner. 
Hence our purpose here is to derive the Hamiltonian (\ref{e:HN}) by 
starting in the Lorentzian formulation with the coordinates $x^{\mu}=(x^0,x^1)$ and the metric $\eta_{\mu\nu}=\mathrm{diag}(-1,+1)$\,.  

\medskip

The deformed Lagrangian is given by 
\begin{multline}
\cal{L}=
	\frac{1}{2\alpha}\biggl(1-
	\sqrt{
		1 - 2\alpha\Bigl[\qty(\del_0\vec{\phi})^2-\qty(\del_1\vec{\phi})^2\Bigr]-4\alpha^2
		\Bigl[\qty(\del_0{\vec{\phi}})^2 \qty(\del_1\vec{\phi})^2-\qty(\del_0 \vec{\phi}\cdot \del_1 \vec{\phi})^2\Bigr]
		}\biggr)\,. 
\end{multline}
Then, the conjugate momentum for $\vec{\phi}$ is 
\begin{align}
\vec{\pi}&\coloneqq \pdv{\cal{L}}{\qty(\del_0\vec{\phi})}
\notag \\&
=\frac{\del_0\vec{\phi} +2\alpha\qty[
\qty(\del_1\vec{\phi})^2\del_0\vec{\phi}
-\qty(\del_0\vec{\phi}\cdot \del_1\vec{\phi})\del_1\vec{\phi}
]}{
\sqrt{
		1 - 2\alpha\qty(\qty(\del_0\vec{\phi})^2-\qty(\del_1\vec{\phi})^2)-4\alpha^2
		\qty(\qty(\del_0{\vec{\phi}})^2 \qty(\del_1\vec{\phi})^2-\qty(\del_0 \vec{\phi}\cdot \del_1 \vec{\phi})^2)
		}
}\,. 
\end{align}
By solving it with respect to $\del_0 \vec{\phi}$\,, we obtain 
\begin{align}
\del_0{\vec{\phi}}
=\frac{
	\vec{\pi} +2\alpha\qty(\vec{\pi}\cdot\del_1\vec{\phi})\del_1\vec{\phi}
}{
\sqrt{
	1 +2\alpha\qty(\del_1\vec{\phi})^2 +2\alpha\vec{\pi}^2+4\alpha^2\qty(\vec{\pi}\cdot\del_1\vec{\phi})^2
	}
}
\end{align}
and the Hamiltonian (\ref{e:HN}) is expressed as 
\begin{align}
\cal{H}&\coloneqq \vec{\pi} \cdot \del_0\vec{\phi}-\cal{L}\eval{}_{\del_0\vec{\phi}=\del_0\vec{\phi}\qty(\vec{\phi},\vec{\pi})}
\nonumber\\ \notag
&=\frac{1}{2\alpha}\qty[
-1+
\sqrt{
		1+2\alpha\qty(\del_1\vec{\phi})^2+2\alpha\vec{\pi}^2+4\alpha^2\qty(\vec{\pi}\cdot\del_1\vec{\phi})^2
	}
]\,.
\end{align}

Let us discuss the relation between the signature of $\alpha$ and stability of the system.
When $\alpha$ is positive, $\cal{H}$ is real for any $\del_1 \vp$ and $\vpi$, and is bounded from below. 
On the other hand, when $\alpha$ is negative, $\cal{H}$ may be complex 
because the inside of the square root becomes negative for some configurations. 
Even for the configurations for which $\cal{H}$ is real, $\cal{H}$ has no lower bound and the system becomes unstable.
Therefore, $\alpha$ should be positive in order to obtain a consistent quantum theory.

\medskip

So far, we have discussed at classical level. 
Let us make a comment on the quantization of the Hamiltonian in the operator formalism. Since the Hamiltonian includes $\partial_1\vec{\phi}$ and $\vec{\pi}$\,, one may worry about the ordering problem. However, these commute with each other at the identical point because we can show the following relation: 
\begin{align}
\comm{\del_1\phi_i(x^0,x^1)}{\pi_j(x^0,y^1)}&=\lim_{y^1\to x^1}\comm{\del_1\phi_i (x^0,x^1)}{\pi_j(x^0,y^1)} \notag \\
&=\lim_{y^1\to x^1}\delta_{ij}\del_1\delta(x^1-y^1). 
\end{align}
Since $\partial_1\delta(x^1-y^1)$ is an odd distribution, the limit must vanish. Therefore there is no ordering problem. Note that this is a special property in the {\it massless} case. If we consider the massive case, $\vec{\phi}$ appears in the Hamiltonian and it causes the ordering problem. It is also obvious that the Hamiltonian is hermite and hence one may use the path-integral formalism as usual.

\section{Calculation of the effective potential}
\label{app:CalcVeff}
In this appendix, we compute the effective potential on an infinitely-long cylinder with the circumference $\beta$\,. 

\medskip

For later use, we first evaluate the following quantity:
\begin{align}
\tr \log(-\del_a c^{ab} \del_b)\,, 
\label{e:trlog_eval}
\end{align}
where $c^{ab}$ is a constant symmetric tensor which satisfies $\det c^{ab}>0$. Let us employ a combined symbol for the integral and sum over the momentum space:
\begin{align}
\frac{1}{\mathrm{Vol}}\,\mathrm{tr}f&=\frac{1}{\mathrm{Vol}}\int_{-\infty}^{\infty} \frac{dk_1}{2\pi} \sum_{n\in \mathbb{Z}} \frac{1}{\beta}\left<k_1,k_2\right|f\left|k_1,k_2\right>\notag\\
&\equiv\int_k\!\!\!\!\!\!\!\!\sum\,f\qty(k_1,k_2)\,, 
\end{align}
where $k_2$ denotes $2\pi n/\beta$ and ``\,Vol\,'' is the coordinate space volume.  The quantity (\ref{e:trlog_eval}) is then written as 
\begin{align}
\frac{1}{\mathrm{Vol}}\mathrm{tr} \log(-\del_a c^{ab} \del_b)&=
\int_k\!\!\!\!\!\!\!\!\sum\, \log(c^{ab}k_a k_b)
\notag \\
&=\int_k\!\!\!\!\!\!\!\!\sum\, \log\qty(c^{11}k_1^{\,2}+2c^{12}k_1k_2+c^{22}k_2^{\,2}) 
\notag \\&=\int_k\!\!\!\!\!\!\!\!\sum\, \log\qty(c^{11})+\int_k\!\!\!\!\!\!\!\!\sum\,\log(k_1^{\,2}+2\frac{c^{12}}{c^{11}}k_1k_2+\frac{c^{22}}{c^{11}}k_2^{\,2})
\label{e:trlog_eval2}
\end{align}
The first term is a constant and is to be subtracted by choosing an appropriate measure of the path integral as in the ordinary field theories. 
On the other hand, the second term in the last expression of  (\ref{e:trlog_eval2}) can be evaluated just as the Casimir energy on a cylinder. 
By introducing a new quantity
\begin{align}
R\coloneqq \ifrac{\abs{c^{11}}\beta}{\sqrt{\det c^{ab}}}\,,
\end{align}
it can be evaluated as 
\begin{align}
\int_k\!\!\!\!\!\!\!\!\sum\,\log(k_1^{\,2}+2\frac{c^{12}}{c^{11}}k_1k_2+\frac{c^{22}}{c^{11}}k_2^{\,2}) 
&=\int_k\!\!\!\!\!\!\!\!\sum\, \log\qty[\qty(k_1+\frac{c^{12}}{c^{11}}k_2)^2+\frac{\det c^{ab}}{(c^{11})^2}k_2^{\,2}]\notag\\
&=\frac{\abs{c^{11}}}{\sqrt{\det c^{ab}}}\int_{-\infty}^{\infty} \frac{dk_1}{2\pi}\frac{1}{R}\sum_{n \in \mathbb{Z}} \log\qty[k_1^2+\qty(\frac{2\pi}{R} n)^2]\notag\\
&\sim\frac{\abs{c^{11}}}{\sqrt{\det c^{ab}}}\qty(-\frac{\pi}{3R^2})\notag\\
&=-\frac{\sqrt{\det c^{ab}}}{\abs{c^{11}}}\frac{\pi}{3\beta^2}\,.
\label{e:cyl_f}
\end{align}
We have again subtracted a divergent constant, which also corresponds to choosing the measure of the path integral or zeta-function regularization. 

\medskip
In summary, we have obtained the following formula:
\begin{align}
    \tr \log(-\del_a c^{ab} \del_b) = \mathrm{Vol}\times \qty(-\frac{\sqrt{\det c^{ab}}}{\abs{c^{11}}}\frac{\pi}{3\beta^2})\,.
    \label{formula}
\end{align}
Using the above formula (\ref{formula}), we can readily evaluate \siki{e:se_g}. When the auxiliary fields $\lambda$\,, $\sigma$ and $\rho$ are constant, the first term in (\ref{e:se_g}) can be computed as
\begin{align}
&
\frac{1}{2}\tr \log \qty[\lambda \del_1 \lambda \del_1-\frac{1}{4}(\del_2+2\del_1 \lambda \sigma )(\del_2+2\lambda \sigma \del_1)]
\notag \\
 =& \frac{1}{2}\int\!\! d^2x\, \int_k\!\!\!\!\!\!\!\!\sum\,
\log \qty[-\lambda^2 (1-\sigma^2)k_1^2 +\lambda\sigma k_2k_1 +\frac{1}{4}k_2^2]
\notag \\ =& \int\!\! d^2x\, \qty(-\frac{\pi }{6\beta^2}\frac{1}{2\abs{\lambda}\abs{1-\sigma^2}})\,.
\end{align}
In the above computation, we have identified the components of $c^{ab}$ as 
\begin{eqnarray}
c^{11}=-\lambda^2(1-\sigma^2)\,, \quad  c^{21}=c^{12}=\lambda\sigma/2\,, 
\quad c^{22}=1/4\,.
\end{eqnarray}
Thus we have derived the effective potential (\ref{e:Veff}).

\section{Revisiting the derivation of the Burgers equation}
\label{app:ExactBurgers}

In this Appendix, we will revisit the derivation of the Burgers equation in the operator formalism under some conditions.  

\subsection{Response under parameter variation}

Here, we will discuss a general system defined on a $(d-1)$ dimensional space with the metric 
\begin{align}
g_{\mu\nu} dx^\mu dx^\nu=-dt^2+h_{ij}(x)dx^idx^j
\end{align}
and with some parameter $\gamma$\,. Suppose here that $\gamma$ is irrelevant to the space volume $\int d^{d-1}\vb*{x}$\,. 

\medskip

We will show that the following relation holds:
\begin{align}
\dv{\gamma }H
&=-\dv{\gamma}L
	\eval{}_{\del_0\vec{\phi}=\del_0\vec{\phi}\qty(\vec{\phi},\vec{\Pi})}\,,
	\label{e:DerHL}\\
L&=\int\!\! d^{d-1}\vb*{x}\,\sqrt{h}~\mathcal{L}^{(\gamma,h)}\,.
\end{align}
Here $\vec{\Pi}$ denotes the conjugate momentum for $\vp$ that is defined as a scalar density: 
\begin{align}
\label{e:defPi}
\vec{\Pi} \coloneqq\pdv{(\del_0 \vp)} \qty(\sqrt{h}\cal{L}^{(\gamma,h)})\,,
\end{align}
and $\del_0\vp(\vp,\vec{\Pi})$ is the solution to \siki{e:defPi} as a function of $\vp$ and $\vec{\Pi}$\,. 
The latter implicitly depends on $\gamma$ and $h_{ij}$\,. The associated classical Poisson structure and Hamiltonian are defined respectively as 
\begin{align}
\{\phi_n(t,\vb*{x}), \Pi_m(t,\vb*{y})\}&=\delta_{nm}\delta^{d-1}(\vb*{x}-\vb*{y})\,,
\label{e:Poisson}\\
H&\coloneqq \int d^{d-1}\vb*{x}\qty(\vec{\Pi}\cdot \del_0 \vp-\sqrt{h}\cal{L}^{(\gamma,h)})\eval{}_{\del_0\vp=\del_0\vp(\vp,\vec{\Pi})}\,.
\end{align}
Note that \siki{e:Poisson} is independent of $h_{ij}(x)$\,.

\medskip

Then we can easily prove the relation (\ref{e:DerHL}) like 
\begin{align}
\dv{\gamma}H
&=\int\!\! d^{d-1}\vb*{x}\, \dv{\gamma}\qty[
	\vec{\Pi} \cdot \del_0\vec{\phi}-\sqrt{h}\cal{L}^{(\gamma,h)}
	]\eval{}_{\del_0\vec{\phi}=\del_0\vec{\phi}\qty(\vec{\phi},\vec{\Pi})}
\notag\\&=\int\!\! d^{d-1}\vb*{x}\, \biggl(
	\vec{\Pi} \cdot \dv{(\del_0\vec{\phi})}{\gamma}
		-\biggl[
			\dv{(\del_0\vec{\phi})}{\gamma}\cdot \pdv{(\del_0 \vp)}+\pdv{\gamma}\biggr]
		\sqrt{h}\cal{L}^{(\gamma,h)}
		\biggr)
	\bigg|_{\del_0\vec{\phi}=\del_0\vec{\phi}\qty(\vec{\phi},\vec{\Pi})}
\notag\\&=\int\!\! d^{d-1}\vb*{x}\, \biggl(
	\vec{\Pi} \cdot \dv{(\del_0\vec{\phi})}{\gamma}
		-\dv{(\del_0\vec{\phi})}{\gamma}\cdot \vec{\Pi}-\pdv{\gamma}
		(\sqrt{h}\cal{L}^{(\gamma,h)})\biggr)
	\bigg|_{\del_0\vec{\phi}=\del_0\vec{\phi}\qty(\vec{\phi},\vec{\Pi})}
\notag\\&=-\int\!\! d^{d-1}\vb*{x}\, \pdv{\gamma}(\sqrt{h}\cal{L}^{(\gamma,h)})
	\eval{}_{\del_0\vec{\phi}=\del_0\vec{\phi}\qty(\vec{\phi},\vec{\Pi})}
\notag\\&=-\dv{\gamma}L
	\eval{}_{\del_0\vec{\phi}=\del_0\vec{\phi}\qty(\vec{\phi},\vec{\Pi})}\,.
\label{e:HLproof}
\end{align}
Therefore, in the context of $\ttb$-deformation, we find that 
\aln{
\dv{H}{\alpha}=-\dv{L}{\alpha}\eval{}_{\del_0\vec{\phi}=\del_0\vec{\phi}\qty(\vec{\phi},\vec{\Pi})} 
\label{e:alphaHL}
}
at classical level.

\medskip

It is remarkable that one may also consider a variation with $h_{ij}(x)$ as above. It results in the following relation:
\aln{
\delta H &= - \delta L	\Big|_{\del_0\vec{\phi}=\del_0\vec{\phi}\qty(\vec{\phi}\,,\vec{\Pi})}\notag\\
&=-\frac{1}{2}\int\!\! d^{d-1}x\,\sqrt{h}\,T^{ij}\delta h_{ij}\,, 
\label{e:HLT}
}
where $T^{ij}(x)$ is the energy-momentum tensor. 

\medskip

Finally, note that the two relations (\ref{e:DerHL}) and (\ref{e:HLT}) can easily be promoted to the quantum level in some special cases 
where the Hamiltonian and Lagrangian consist of commutative operators $\vec{\Pi}$ and $\partial_1\vec{\phi}$\,, including our present case, a $\ttb$-deformed massless $O(N)$ vector model.
Then one can formally take derivatives with respect to them (in particular, one can use the chain rule in the second term in the second line in \siki{e:HLproof})\,.

\subsection{Derivation of the Burgers equation}
\label{app:DeriveBurgers}

Let us next derive the Burgers equation for a $\ttb$-deformed theory on a cylinder with circumference $R$\,.

\medskip

The coordinate for the circumference direction is represented by $x^1$ with the normalized period: 
\begin{align}
x^1\sim x^1+1\,. 
\end{align}
In turn, the metric $g_{\mu\nu}$ includes the information of the circumference:  
\begin{align}
g_{\mu\nu} &=\diag (-1,R^2)\,,\notag\\
h_{11} &= R^2\,.
\label{e:h11}
\end{align}
Note that in this coordinate system, the $\ttb$-flow equation is written as 
\begin{align}
\pdv{\alpha} \cal{L}^{(\alpha)} =  R^{-2}\det \qty({T^{(\alpha)}_{~\mu\nu}})\,.
\end{align}

\medskip

In the following, we will suppose the conditions that 1) the $\ttb$ operator is well-defined as a local operator, 2) the expectation value of the $\ttb$ operator is factorized and 3) the Hamiltonian consists only of mutually commutative operators. 
Then we will discuss at quantum level. 

\medskip

The quantum version of (\ref{e:alphaHL}) is given by 
\begin{align}
\dv{{\hat{H}}^{(\alpha)}}{\alpha}
&=-\int\!\! dx\, R^{-1} \det\qty({\hat{T}^{(\alpha)}_{~\mu\nu}})\,.
\label{e:ttbflowH}
\end{align}
By taking the expectation value of both sides of (\ref{e:ttbflowH}) in the $n$-th excited state $\ket{n}$\,. Now notice that the following relation holds:
\begin{align}
\ev{\dv{\hat{H}^{(\alpha)}}{\alpha}}_n
=R^{-1}\ev{\det \qty({\hat{T}^{(\alpha)}_{~\mu\nu}})}_n
= - R^{-1}\qty(\ev{\hat{T}_{~00}^{(\alpha)}}_n\ev{\hat{T}_{~11}^{(\alpha)}}_n-\ev{\hat{T}_{~01}^{(\alpha)}}_n^2)\,,
\label{e:mBurgers}
\end{align}
where we have defined $\ev{\cal{O}}_n$ by $\bra{n}\cal{O}\ket{n}$\,.
We have used the translational symmetry of the eigenstates and the factorization property of $\det \hat{T}_{\mu\nu}^{(\alpha)}$  \cite{Zamolodchikov:2004ce}. 

\medskip

Let us evaluate the right-hand side of (\ref{e:mBurgers})\,. 
Here, each state is an eigenstate simultaneously for the Hamiltonian  and the total momentum:
\begin{align}
\hat{H}^{(\alpha)}\ket{n} &=E_n^{(\alpha)}\ket{n} \qquad \qty(\hat{H}^{(\alpha)}\coloneqq \int\!\! dx\, R\,\hat{T}^{(\alpha)}_{~00})\,,
\\\hat{P}^{(\alpha)}\ket{n} &=P_n \ket{n} \qquad\quad  \qty(\hat{P}^{(\alpha)}\coloneqq\int\!\! dx\, \hat{T}^{(\alpha)}_{~01})\,,
\end{align}
and hence we obtain 
\aln{
\ev{\hat{T}^{(\alpha)}_{~00}}_n=\frac{E_n^{(\alpha)}}{R}\,, \qquad  \ev{\hat{T}^{(\alpha)}_{~01}}_n=\ev{\hat{T}^{(\alpha)}_{~10}}_n=P_n\,.
\label{e:evT00T01}
}
Note here that $P_n$ is not changed under the $\ttb$-deformation because it is determined by the winding number. 
As for $\ev{\hat{T}^{(\alpha)}_{~11}}_n$\,, the differentiation of the Hamiltonian by $R$ yields to 
\begin{align}
\dv{H^{(\alpha)}}{R}
&=\dv{(R^{2})}{R}\dv{\hat{H}^{(\alpha)}}{(R^{2})}\notag\\
&=2R \dv{\hat{H}^{(\alpha)}}{h_{11}}\notag\\
&=-2R \int_0^1\!\! dx^1\, \frac{R}{2}\hat{T}^{(\alpha)11}\notag\\
&=-R^{-2}\int_0^1\!\! dx^1\, \hat{T}^{(\alpha)}_{~11}\,.
\label{e:DerH}
\end{align}
By taking the expectation value of both sides, we get\footnote{
We have utilized the Hellmann-Feynman theorem. 
}
\aln{
\ev{\hat{T}^{(\alpha)}_{~11}}=-R^2\dv{E_n^{(\alpha)}}{R}\,.
\label{e:evT11}
}

\medskip

On the other hand, the left-hand side of (\ref{e:mBurgers}) can be obviously evaluated as  
\aln{
\ev{\dv{\hat{H}^{(\alpha)}}{\alpha}}_n=\pdv{E_n^{(\alpha)}}{\alpha}\,.
\label{e:evdHdalpha}
}
By substituting (\ref{e:evT00T01}), (\ref{e:evT11}) and (\ref{e:evdHdalpha}) into (\ref{e:mBurgers}), we obtain the Burgers equation:
\begin{align}
\pdv{E_n^{(\alpha)}}{\alpha}
&= E^{(\alpha)}_n\pdv{E_n^{(\alpha)}}{R}
+ R^{-1}\qty(P_n)^2.
\label{e:Burgers}
\end{align}

\end{document}